\documentstyle[preprint,prc,aps]{revtex}
\begin{document}
\title{THE SHORT-RANGE BARYON-BARYON INTERACTION IN A CHIRAL CONSTITUENT QUARK MODEL}
\author{Fl. Stancu }
\address{Universit\'{e} de Li\`ege, Institut de Physique B.5, Sart Tilman,
B-4000 Li\`ege 1, Belgium}
\author{L. Ya. Glozman}
\address{High Energy Accelerator Research Organization (KEK), 
Tanashi Branch, Tanashi, Tokyo 188-8501, Japan}
\date{\today}
\maketitle
\begin{abstract}
The previous analysis of the short-range $NN$ repulsion
originating from the Goldstone boson exchange hyperfine interaction
between constituent quarks is revisited. 
We study in which respects the repulsion depends on 
the radial form of the spin-spin quark-quark force. We show that while
the radial form affects the structure of the 6Q wave function,
the short-range repulsion in the $NN$ system persists in any case. We extend
our analysis to other $YN$ and $YY$ (flavor octet-octet) systems and
demonstrate that the flavor-spin hyperfine interaction implies a
short-range repulsion in these B=2 systems as well.
\end{abstract}

\bigskip
\bigskip

\section{Introduction}

An intricate question of the intermediate energy nuclear
physics is about the origin of the short-range repulsion in the  
nucleon-nucleon ($NN$) system or, more generally, in baryon-baryon
systems. By now it is clear that the mechanism 
describing the $NN$ interaction should be related
with the QCD dynamics responsible for the low-energy properties
of the nucleon. It is also now evident that the most important
QCD phenomenon in this case is the spontaneous breaking of chiral
symmetry, which implies that at momenta below the chiral
symmetry breaking scale the relevant quasiparticle degrees
of freedom are constituent quarks and pseudoscalar mesons,
which are Goldstone bosons of the broken chiral symmetry.
Assuming that in this regime the dominant interaction
between confined constituent quarks is due to Goldstone
boson exchange (GBE) one can understand the structure of the whole
low-lying baryon spectrum \cite{GR96,GPVW}. A GBE interaction
between constituent quarks is a natural interpretation of
the t-channel iterations of point-like gluonic interactions
between quarks which are responsible for dynamical breaking
of chiral symmetry in QCD vacuum \cite{GV}.\\

In a previous work \cite{STPGL} it was shown that the short-range
part of the flavor-dependent spin-spin force between constituent
quarks, which is due to GBE, and which is reinforced by the
short-range part of vector-meson exchange \cite{G} (correlated
two-GBE \cite{RB}), induces a strong short-range repulsion
in the $NN$ system. The same interaction also implies a strong
short-range repulsion in the 6Q system with ``$H$-particle''
quantum numbers \cite{H}, suggesting that a very existence of a deeply
bound $H$-particle is impossible within the given picture.
In the present work we extend our analysis to other flavor
octet-octet B=2 systems and show that the short-range
repulsion persists as a general case.

In section II we revisit the $NN$ interaction and discuss in 
which respects the predictions of \cite{STPGL} depend on the
radial form of the short-range flavor-spin interaction.
Section III is devoted to the classification of 
6Q states relevant for the hyperon-nucleon 
($YN$) and hyperon-hyperon ($YY$) systems and to a qualitative estimate
of the short-range repulsion appearing in these systems.

\section{The $NN$ interaction at short range - revisited}
The results of Ref. \cite{STPGL} are based on the  
flavor-spin hyperfine interaction between two constituent quarks
$i$ and $j$
which has a short-range part of the form
\cite{GR96}

\begin{equation}
- \lambda_{i}^{F} \cdot \lambda_{j}^{F}
\vec{\sigma}_i \cdot \vec{\sigma}_j ,
\label{opFS}
\end{equation}

\noindent
where $\lambda^{F}$ 
are the flavor Gell-Mann  matrices and an implicit summation over $F$
=1,2,...8 is understood. The operator (\ref{opFS}) represents
the short-range part of the Goldstone boson exchange interaction \cite{GR96}.
Two correlated Goldstone bosons (vector meson exchange)
enhance the effect of the short-range part of the one-boson exchange
interaction,
as shown in \cite{G,RB}, so it can be incorporated in (\ref{opFS}) as well. 
In Ref.  \cite{STPGL} it was shown that such a flavor-spin interaction 
leads to a short-range repulsion in the $NN$ interaction when 
the latter is treated as a 6Q system. 
This repulsion results from the fact that the 
energy of the most favourable compact 6Q configuration,
$s^4p^2 [42]_O [51]_{FS}$, is much above the two-nucleon threshold.
Here and below $[f]_O, [f]_{FS}$, and $[f]_F$ denote Young
diagrams specifying the permutational orbital (O), $SU(6)_{FS}$
(FS), and $SU(3)_F$ (F) symmetries and it is always assumed that
the center-of-mass motion is removed from the shell model
configurations.
In \cite{STPGL} the nonrelativistic Hamiltonian with the 
parametrization of the hyperfine interaction of Ref. 
\cite{GPP96} has been used.
That parametrization, being successful for baryon spectra,
may lead, however, to some undesirable effects in the $NN$ system
because it contains the shift parameter $r_0 = 0.43$ fm of the
short-range hyperfine interaction from the origin:

\begin{equation}4\pi \delta(\vec r_{ij}) \Rightarrow \frac {4}{\sqrt {\pi}}
\alpha^3 \exp(-\alpha^2(r-r_0)^2). \label{CONTACT} \end{equation}

\noindent
This shift enhances the quark-quark matrix elements
with $1p$ relative motion and affects the coupling between 
the symmetry states chosen for the 
the diagonalization of the Hamiltonian. Here we try to find out to which
extent the results of Ref. \cite{STPGL} are modified
when one takes $r_0$ = 0. Actually 
the baryon spectra can be described as well
without such a shift with a properly chosen
parametrization \cite{GPVW}.

To estimate the strength of the short-range $NN$ interaction 
we use the adiabatic (Born-Oppenheimer) 
approximation

\begin{equation}
V_{NN}(R) = \langle H \rangle_R - \langle H \rangle_{\infty},
\label{born}
\end{equation}

\noindent
where $R$ is a generator coordinate, defined as the distance
between two harmonic oscillator wells, 
each associated asymptotically to a $3Q$ 
cluster. 
In Eq. (\ref{born}) 
$\langle H \rangle _R$ is the lowest eigenvalue 
resulting from the diagonalization  
of the Hamiltonian at fixed $R$. Presently we  are interested in the
effective potential at $R$ = 0 only. In this case in the S-wave
 relative motion only the $s^6 [6]_O$ and $s^4p^2[42]_O$ 
shell-model configurations
are allowed \cite{HAR}. So to estimate the strength
of the repulsion we diagonalize the Hamiltonian in the
basis of the following four most important configurations 
\cite{STPGL}:

\begin{equation}
\begin{array}{ccc}
|1> &=&| s^6 [6]_O [33]_{FS} > \\
|2> &=&| s^4p^2 [42]_O [33]_{FS} > \\
|3> &=&| s^4p^2 [42]_O [51]_{FS} > \\
|4> &=&| s^4p^2 [42]_O [411]_{FS} >. \\
\end{array}
\label{basis}
\end{equation}

The Hamiltonian and the confinement potential are taken
from the ref. \cite{GPP96}. Here we modify the radial
dependence (see eqs. (17)-(18) of ref. \cite{STPGL}) of
the meson-exchange potential as follows:

\medskip
(i) we drop the long-range Yukawa part 
${\mu_\gamma^2~{\exp(-\mu_\gamma r_{ij})}~/ r_{ij}}$ of
the potential, as it contributes very little  
at short-range in the 6Q system;

\medskip
(ii) in the short-range part of the
hyperfine interaction we put $r_0$ = 0;

\medskip
(iii) we keep the ratio of the singlet ($g_0$) to octet ($g_8$) coupling
constants as in Refs. \cite{STPGL,GPP96} but readjust the
absolute value of each coupling constant in order to reproduce
the $\Delta - N$ mass splitting keeping in mind the modifications (i) and (ii).

\medskip
The root-mean-square matter radius $\beta$ of the $s^3$ nucleon and $\Delta$,
which coincides with the harmonic oscillator parameter in the 6Q basis,
is obtained from the nucleon stability condition

\begin{equation}
 \frac {\partial}{\partial \beta} \langle N | H | N \rangle = 0 .
\label{STAB} \end{equation}

This procedure gives $g_8^2/4\pi = 2.11$, which should be regarded
as an effective coupling constant simulating the combined effect of
both the pseudoscalar- and the vector-meson-like exchange short-range
interaction (\ref{opFS}) with a nonrelativistic $s^3$ ansatz
for the baryon. The minimum of $m_N$ = $\langle N|H|N \rangle$ = 1.324 GeV 
is achieved
at $\beta$=0.373 fm. The absolute value of the ``nucleon mass'' $m_N$
is unimportant in the present context as it identically 
cancels out in (\ref{born}). The calculated $m_N$ can 
also be shifted to a physical value
by simply adding a constant contribution to the effective
confining interaction.

The results of the diagonalization for the $^3S_1$ and $^1S_0$
NN partial waves are shown in Tables I and II, which should be
compared with Tables II and III of Ref. \cite{STPGL} or 
more precisely to Tables V and VI of Ref. \cite{BS98} because the 
latter tables show the diagonalization results of a 4 $\times$ 4 matrix, 
as here. In Ref.  \cite{BS98} the 5th basis vector used in \cite{STPGL}
has also been removed because it has no correspondence in the molecular
orbital basis employed there.

We see that with the present parametrization of the short-range
QQ interaction the $s^4p^2[51]_{FS}$ configuration is again the
lowest one with an energy of roughly 1 GeV below the energy
of the orbitally unexcited configuration $s^6[33]_{FS}$.
In this sense the conclusion of Ref. \cite{STPGL}
is reconfirmed. The lowest eigenvalue is about 1.4 GeV above
the $2m_N$ threshold, which shows that the strong short-range
repulsion in the $NN$ system persists with this new parametrization.
However, the  numerical values of the off-diagonal matrix
elements are now different compared to those 
obtained with the parametrization of Refs. \cite{STPGL} or \cite{BS98}.
As a consequence the amplitude of the configuration $s^4p^2[51]_{FS}$
is somewhat smaller and one finds more mixing among  
the configurations $|1\rangle - |4\rangle$ in the lowest
state eigenvector, in contrast to Refs. \cite{STPGL} or  \cite{BS98}.
But the configuration $s^4p^2[51]_{FS}$ remains dominant 
in the lowest state and it can induce additional effective
repulsion as discussed in \cite{STPGL} (see also \cite{NST,OY}).
While the
mixing does not affect the conclusion about the 
repulsive core, it is important for the behaviour of the
6Q wave function at short range. When one projects this wave function
on the $NN$ channel according to the procedure described in  \cite{STPGL} 
the node, predicted there  (see Fig. 1), nearly disappears. 
Now the behaviour of the projection
is similar to the $NN$ wave function obtained with usual 
repulsive core potentials. This behaviour comes from the
destructive interference of the excited $s^4p^2$ and
nonexcited $s^6$ configurations. Such a destructive interference
has been observed earlier in other microscopical models
\cite{OB,GLBK} and can be considered as a substantiation of
the repulsive core in $NN$ potentials. However, contrary to any
simple $NN$ potential model, the 6Q wave function is much
richer and contains not only the $NN$ component, but also a variety of 
other components such as  $NN^*, N^*N^*, \Delta\Delta, \Delta\Delta^*, ...$
\cite{GLKU}.

Next we adress the question to which extent the height
of the repulsion in the $NN$ system is sensitive to the
contribution of the flavor-singlet $\eta'$-like exchange.
The $\eta'$-like exchange tends to decrease the
$\Delta - N$ splitting, while the $\pi$-like exchange
works just in opposite direction. It means that in reproducing the
$\Delta - N$ splitting the octet coupling constant
will become smaller once the  $\eta'$-exchange interaction
is dropped.  We
take the extreme limit $g_0^2/4\pi=0$ and repeat the
steps (i)-(iii) from above. One obtains $\beta$ = 0.522 fm,
$ g_8^2/4\pi=1.29$ and $m_N = 1.4657$ GeV. The results
for the $NN$ system are given in Tables III and IV. Comparing
them with those of Tables I and II we conclude that
the height of the repulsive core is essentially
smaller than before and 
the structure of the ground state eigenvector is changed.
In the present case the configurations
$s^6[33]_{FS}$ and $s^4p^2[51]_{FS}$ become approximately
degenerate and still about 900 MeV above the $2m_N$ threshold.
The lowest eigenvalue is about 600 MeV above the threshold,
showing that the strength of the repulsion is reduced
compared to the previous case, but still important.

The calculation of the $NN$ interaction at short range
within the adiabatic approximation above should be
taken with some caution, however.
In fact it represents only the diagonal kernel 
of a dynamical treatment such as the resonating group method (RGM).
So an ultimate conclusion could only be drawn from the
behavoiur of the $NN$ phase shifts calculated
beyond the adiabatic approximation. Such phase shifts,
obtained within an extended resonating group method,
do indicate the presence of 
a very strong repulsion even in the case without
any $\eta'$-exchange interaction \cite{SG}.
  
\section{ Short-range 6Q configurations in hyperon-nucleon and
hyperon-hyperon systems}

In this section we discuss the issue whether or not the short-range
repulsion in the $NN$ system, implied by the flavor-spin hyperfine
interaction (\ref{opFS}), persists in other baryon-baryon
(flavor octet-octet) systems. We first construct the lowest possible
symmetry states $|[f]_O [f]_F [f]_{FS}>$ compatible with
the asymptotic two-baryon channels. For simplicity we
consider the $SU(3)_F$ limit (see Eq. (\ref{HAM}) below).

Assuming that the orbital wave function of 
any of the octet baryons is described by the $[3]_O$
permutational symmetry, only two states $[6]_O$ and
$[42]_O$ are allowed in the S-wave relative motion of a two-baryon
system \cite{HAR,NST}. Applying the inner product rules of the
symmetric group both for $[f]_O \times [f]_C$, where
$[f]_C = [222]_C$ is fixed by the color-singlet nature of
the two-baryon system, and  $[f]_F\times [f]_S$, where
$[f]_S$ is fixed by the total spin $S$ of the 6Q system,
one arrives at the symmetry states
listed in Tables V and VI. We recall that for a given $[f]_S$
of $SU(2)$ where $f_1$ and
$f_2$ represent the number of boxes in the first and second rows of the
Young diagram $[f]_S$ respectively, with $f_1 + f_2 = 6$ in the present
case, the spin is given by $S = 1/2 (f_1-f_2)$.

When one considers a schematic model \cite{GR96}, where the interaction
Hamiltonian is approximated as

\begin{equation}
H_\chi = - C_\chi \sum_{i < j} \lambda_{i}^{F} \cdot \lambda_{j}^{F}
\vec{\sigma}_i \cdot \vec{\sigma}_j ,
\label{HAM}
\end{equation}

\noindent 
and the constant $C_\chi$ = 29.3 MeV is extracted from the phenomenological
$\Delta - N$ splitting, then the expectation
value of the interaction (\ref{HAM}) for all symmetry states
in Tables V and VI can be evaluated through the Casimir operator
eigenvalues, see Appendix A of ref. \cite{STPGL}. The results
are given in Tables V and VI in units of $C_\chi$. One can immediately
conclude from these Tables that far the lowest state  
is $| s^4p^2 [42]_O [321]_F [51]_{FS}>$.
This state is not allowed in the $NN$ system. Here it appears
due to the presence of three distinct flavors $u$, $d$ and $s$. 
Its contribution is lower than that
of the vector $|3>$ of the list (\ref{basis})
by about 10 $C_{\chi}$ units 
(recall that in the $NN$ case the flavor symmetry is $[42]_F$ for $S = 0$ 
S-wave and $[33]_F$ for $S = 1$ S-wave). By itself the 
$| s^4p^2 [42]_O [321]_F [51]_{FS}>$ state
guarantees that a strong effective repulsion will persist
in the $YN$ and $YY$ systems, again related to a specific
symmetry structure of the type $s^4p^2 [51]_{FS}$.
The ground state configuration
$s^6[6]_O$ becomes even more ``forbidden'' by dynamics than
in the $NN$ system (i.e. the weight of $s^6[6]_O$ should be
expected to become smaller). This effect, however, cannot be obtained
within the simple approximation considered below
where only the expectation value of the
$| s^4p^2 [42]_O [321]_F [51]_{FS}>$ state
is calculated.\\

To have a rough qualitative idea about the strength of the
interaction at short range in $YY$ and $YN$ systems we thus
calculate the diagonal matrix element

\begin{equation}
<s^4p^2 [42]_O [321]_F [51]_{FS} | H_0 + H_{conf} + H_\chi |
s^4p^2 [42]_O [321]_F [51]_{FS} >
\label{ME}
\end{equation}

\noindent 
and compare it with the two-baryon threshold. With the
coupling constant $g_8^2/{4\pi} = 2.11$, and the ratio
$g_0^2/g_8^2$, fixed in the previous section, we now use
a harmonic oscillator parameter $\beta = 0.403 $ fm,
which provides an equilibrium value for $\Lambda$, as
it follows from Table VII.

In all cases we describe the kinetic energy 
of a $6Q$ system in a simple way
\begin{equation}
<s^4p^2 | H_0 | s^4p^2 > = \frac{19}{4}\hbar\omega,
\end{equation}
\begin{equation}
\hbar\omega = \frac{\hbar^2}{m_{ave}}\beta^2.
\end{equation}
where $ m_{ave} $ is an average quark mass defined for each system
and the center-of-mass motion is removed.
For example the $\Lambda\Lambda$ system has an average mass
$m_{ave} = (4 m + 2 m_s )/6$, where $m$ = 0.340 GeV and $m_s$ = 0.440 GeV.
\par

All the contributions from $H_\chi$ and $H_{conf}$ are calculated
with the help of the fractional parentage technique, similar
to \cite{STPGL,H} and described in detail in Ref. \cite{ST96}.
The $SU(3)$ Clebsch-Gordan coefficients for the flavor part
of the wave function are taken from Ref. \cite{HECHT}.
By this technique one can  reduce the six-quark matrix elements
to linear combinations of two-quark matrix elements which allow
immediate integration in the spin-flavor space by use of
Eq. (3.3) of Ref. \cite{GR96}. In particular, for the confinement
part $H_{conf}$ the orbital matrix elements
can be easily calculated analytically. 
This gives 
\begin{equation}
\langle H_{conf}\rangle = \frac{71}{6} \sqrt{\frac{2}{\pi}} C \beta.
\end{equation}   
where $C$ is the string tension taken from \cite{GPP96} and $\beta$
has been specified above. 
In an analogue way the confinement energy of a ground state baryon is
\begin{equation}
\langle H_{conf}\rangle = 6\sqrt{\frac{2}{\pi}} C \beta.
\end{equation}
Thus the difference between the $6Q$ and two times the $3Q$ confinement energy 
is $-\sqrt{\frac{2}{\pi}} {\frac{C \beta}{6}}$. This gives 
about $-5$ MeV, which proves that the confinement contribution
nearly cancels out in the baryon-baryon potential in the
present approximation.

Results for $\langle H_0 \rangle$, $\langle H_{\chi} \rangle$
and $\langle H \rangle$
are exhibited in Tables VIII and IX for $S$ = 0 and $S$ = 1 repectively.
We define the separation energy (last column) of either table as the
difference between $\langle H \rangle$ and the lowest threshold
two-baryon mass, calculated with the same hamiltonian,
associated to a given $YI$. The lowest thresholds are
$\Lambda\Lambda$ , $\Sigma\Sigma$ , $ N\Lambda$ and
$\Lambda\Xi$
for $YI$ = 00, 01, 1 1/2 and -1 1/2 respectively.
The separation energy is roughly interpreted as the value taken by 
the baryon-baryon interaction potential  at zero separation distance 
in the Born-Oppenheimer
approximation. The last column of
Tables VIII and IX indicates
repulsion of the order of 1 GeV in all cases. When one
uses a Hamiltonian without $\eta'$-exchange, like at the end
of the above section, then the repulsion is weakened by roughly
200 - 300 MeV.

\section{Conclusions}

This paper closes a series of papers \cite{STPGL,H} devoted
to a qualitative study of the short-range baryon-baryon
interactions within a chiral constituent quark model relying
on meson-exchange dynamics. Our results indicate that the
short-range flavor-spin interaction (1) between constituent quarks
implies a strong short-range repulsion in NN and other flavor octet-octet
systems. While we observe a strong repulsive core we cannot insist
on numerical values for this repulsion because of the use of
simplified approximations. The next stage of the study should
invoke more detailed and refined dynamical approximations
as well as the incorporation of  multiple correlated GBE interactions
(scalar and vector meson exchanges) in order to provide a realistic
description of the middle- and long-range physics in baryon-baryon
systems with all the necessary components, like tensor
and spin-orbit forces. 

\section{Acknowledgement}

L.Ya.G. is indebted to the nuclear theory groups of KEK-Tanashi
and Tokyo Institute of Technology for a warm hospitality. His
work is supported by a foreign guestprofessorship program of
the Ministry of Education, Science, Sports and Culture of Japan.

\vspace{0.8cm}
\begin{table}
{\caption[states]{\label{states} Results of the diagonalization  for
IS = (01). Column 1 - basis states, column 2 - diagonal matrix elements (GeV),
columns 3 - eigenvalues (GeV) in increasing order for a 
4 x 4 matrix ,
column 4 - components of the lowest state. 
The results correspond to $\beta$ =
0.373 fm . The diagonal matrix elements  and
the eigenvalues are relative to 2 $m_{N}$= 2.648 GeV.  }}
\begin{tabular}{|c|c|c|c|}
\hline
State & Diag. elem - 2 $m_N$ & Eigenvalues - 2 $m_N$ & Lowest
state amplitudes\\
\hline
$|s^6[6]_O[33]_{FS} >$ & 2.461 & 1.403 & 0.32885\\
\hline
$|s^4p^2[42]_O[33]_{FS} >$ & 3.119 & 1.914 & -0.25317\\
\hline
$|s^4p^2[42]_O[51]_{FS} >$ & 1.513 & 3.192 & 0.90351\\
\hline
$|s^4p^2[42]_O[411]_{FS} >$ & 3.171 & 3.755 & -0.10691\\
\hline
\end{tabular}
\end{table}

\vspace{0.8cm}
\begin{table}
{\caption[states]{\label{states} Results of the diagonalization  for
IS = (10). See legend of Table I. }}
\begin{tabular}{|c|c|c|c|}
\hline
State & Diag. elem - 2 $m_N$ & Eigenvalues - 2 $m_N$ & Lowest
state amplitudes\\
\hline
$|s^6[6]_O[33]_{FS} >$ & 3.108 & 1.890 & 0.32335\\
\hline
$|s^4p^2[42]_O[33]_{FS} >$ & 3.572 & 2.368 & -0.27724\\
\hline
$|s^4p^2[42]_O[51]_{FS} >$ & 2.009 & 3.778 & 0.89853\\
\hline
$|s^4p^2[42]_O[411]_{FS} >$ & 3.743 & 4.397 & 0.10896\\
\hline
\end{tabular}
\end{table}

\begin{table}
{\caption[states]{\label{states} Results of the diagonalization  for
IS = (01) without $\eta'$-like exchange. Column 1 - basis states, column 2 - diagonal matrix elements (GeV),
columns 3 - eigenvalues (GeV) in increasing order for a 
4 x 4 matrix ,
column 4 - components of the lowest state. 
The results correspond to $\beta$ =
0.522 fm . The diagonal matrix elements  and
the eigenvalues are relative to 2 $m_{N}$= 2.9314 GeV.  }}
\begin{tabular}{|c|c|c|c|}
\hline
State & Diag. elem - 2 $m_N$ & Eigenvalues - 2 $m_N$ & Lowest
state amplitudes\\
\hline
$|s^6[6]_O[33]_{FS} >$ & 0.873 & 0.565 & 0.72185\\
\hline
$|s^4p^2[42]_O[33]_{FS} >$ & 1.326 & 1.032 & -0.43487\\
\hline
$|s^4p^2[42]_O[51]_{FS} >$ & 0.912 & 1.416 & 0.53626\\
\hline
$|s^4p^2[42]_O[411]_{FS} >$ & 1.410 & 1.509 & -0.04737\\
\hline
\end{tabular}
\end{table}

\vspace{0.8cm}
\begin{table}
{\caption[states]{\label{states} Results of the diagonalization for
IS = (10) without $\eta'$-like exchange. See legend to the Table III. }}
\begin{tabular}{|c|c|c|c|}
\hline
State & Diag. elem - 2 $m_N$ & Eigenvalues - 2 $m_N$ & Lowest
state amplitudes\\
\hline
$|s^6[6]_O[33]_{FS} >$ & 0.913 & 0.594 & -0.71686\\
\hline
$|s^4p^2[42]_O[33]_{FS} >$ & 1.352 & 1.056 & 0.44305\\
\hline
$|s^4p^2[42]_O[51]_{FS} >$ & 0.942 & 1.4566 & -0.53765\\
\hline
$|s^4p^2[42]_O[411]_{FS} >$ & 1.445 & 1.545 & 0.02746\\
\hline
\end{tabular}
\end{table}

\begin{table}
\caption{Expectation value $\langle H_{\chi}\rangle$ of
the operator (1) for the lowest symmetry states
$\left.{\left|[f]_O[f]_F[f]_{FS}\right.}\right\rangle$ with $S=0$
compatible with asymptotic two baryon channels.}
\begin{tabular}{|l|l|}
\hline
State & $\langle V_{\chi}\rangle/C_{\chi}$\\
\hline
$\left.{\left|\left[6\right]_O\left[222\right]_F\left[33\right]_{FS}\right.}
\right\rangle$ & -24\\
$\left.{\left|\left[42\right]_O\left[321\right]_F\left[51\right]_{FS}\right.}
\right\rangle$ & -42\\
$\left.{\left|\left[42\right]_O\left[42\right]_F\left[51\right]_{FS}\right.}
\right\rangle$ & -32\\
$\left.{\left|\left[42\right]_O\left[222\right]_F\left[33\right]_{FS}\right.}
\right\rangle$ & -24\\
$\left.{\left|\left[42\right]_O\left[321\right]_F\left[33\right]_{FS}\right.}
\right\rangle$ & -18\\
$\left.{\left|\left[42\right]_O\left[42\right]_F\left[33\right]_{FS}\right.}
\right\rangle$ & -8\\
\hline
\end{tabular}
\end{table}

\begin{table}
\caption{Same as Table V but for spin $S=1$.}
\begin{tabular}{|l|l|}
\hline
State & $\langle V_{\chi}\rangle/C_{\chi}$\\
\hline
$\left.{\left|\left[6\right]_O\left[321\right]_F\left[33\right]_{FS}\right.}
\right\rangle$ & -46/3\\
$\left.{\left|\left[6\right]_O\left[33\right]_F\left[33\right]_{FS}\right.}
\right\rangle$ & -28/3\\
$\left.{\left|\left[6\right]_O\left[411\right]_F\left[33\right]_{FS}\right.}
\right\rangle$ & -28/3\\
$\left.{\left|\left[42\right]_O\left[321\right]_F\left[51\right]_{FS}\right.}
\right\rangle$ & -118/3\\
$\left.{\left|\left[42\right]_O\left[33\right]_F\left[51\right]_{FS}\right.}
\right\rangle$ & -100/3\\
$\left.{\left|\left[42\right]_O\left[411\right]_F\left[51\right]_{FS}\right.}
\right\rangle$ & -100/3\\
$\left.{\left|\left[42\right]_O\left[42\right]_F\left[51\right]_{FS}\right.}
\right\rangle$ &-88/3\\
\hline
\end{tabular}
\end{table}

\begin{table}
\caption {Variational solution 
for low lying. $S = 1/2$ baryons as compared to the experimental masses.}
\begin{tabular}{|l|l|l|l|}
\hline
Baryon & Variational & Variational & Experimental\\
& parameter $\beta$ (fm) & solution (GeV) & mass (GeV)\\
\hline
$N$ &0.373 & 1.324 & 0.940\\
\hline
$\Lambda$ & 0.403 & 1.4702 & 1.1156\\
\hline
$\Sigma$ & 0.433 & 1.5050 & 1.1930\\
\hline
$\Xi$ & 0.457 & 1.6120 & 1.3181\\
\hline
\end{tabular}
\end{table}

\begin{table}
\caption {Expectation value (7) of the $6Q$ hamiltonian  for the lowest state
$\left.{\left|\left[42\right]_O\left[321\right]_F\left[51\right]_{FS}\right.}
\right\rangle$ 
with $S=0$. The harmonic oscillator parameter is chosen equal to that of 
$\Lambda$
from Table 1. Column 1 gives the quantum numbers $YI$ compatible with
$\left[321\right]_F$ of $SU(3)$ and the last column the separation energy 
above the lowest the threshold as
implied by Table I.}
\begin{tabular}{|l|l|l|l|l|l|}
\hline
$YI$ & System & $\langle H_0\rangle$ &
$\langle H_{\chi}\rangle$ & $\langle H\rangle$ & Separation energy\\
& & (GeV) & (GeV) & (GeV) & (GeV)\\
\hline
$00$ &$\Lambda\Lambda, \Sigma\Sigma, N\Xi$ & 3.0536 & -1.6056 & 4.0424 & 1.102\\
\hline
$01$ &$\Sigma\Sigma$, $N\Xi$ & 3.0536 & -1.4163 & 4.2317 &  1.222\\
\hline
$1{\frac{1}{2}}$ & $N\Lambda$, $N\Sigma$ & 3.1963 & -2.0200 & 3.6706 & 0.876\\
\hline
$-1{\frac{1}{2}}$ &$\Lambda\Xi$ & 2.9231 & -1.1647 & 4.4528 & 1.371\\
\hline
\end{tabular}
\end{table}

\begin{table}
\caption{Same as Table IV but for $S=1$.}
\begin{tabular}{|l|l|l|l|l|l|}
\hline
$YI$ & System & $\langle H_0\rangle$ &
$\langle H_{\chi}\rangle$ & $\langle H\rangle$ & Separation energy\\
& & (GeV) & (GeV) & (GeV) & (GeV)\\
\hline
$00$ &$\Lambda\Lambda, \Sigma\Sigma, N\Xi$ & 3.0536 & -1.7287 & 3.9193 & 0.979\\
\hline
$01$ &$\Sigma\Sigma$, $N\Xi$ & 3.0536 & -1.4263 & 4.2217 &  1.212\\
\hline
$1{\frac{1}{2}}$ &$N\Lambda$, $N\Sigma$ & 3.1963 & -2.0258 & 3.6649 & 0.871\\
\hline
$-1{\frac{1}{2}}$ &$\Lambda\Xi$ & 2.9231 & -1.2964 & 4.3212 & 1.239\\
\hline
\end{tabular}
\end{table}


\begin{references}
\bibitem{GR96} L. Ya. Glozman and D.O. Riska, Physics Reports {\bf 268}, 263
(1996).
\bibitem{GPVW} L. Ya. Glozman,  W. Plessas, K. Varga
and R.F. Wagenbrunn, Phys. Rev. {\bf D58}, 094030 (1998).
\bibitem{GV} L.Ya. Glozman and K. Varga, hep-ph/9901439.
\bibitem{STPGL} Fl. Stancu, S. Pepin, and L. Ya. Glozman,
Phys. Rev. {\bf C56}, 2779 (1997); [E: {\bf C59} 1219 (1999).
\bibitem{H} Fl. Stancu, S. Pepin, and L. Ya. Glozman,
Phys. Rev. {\bf D57}, 4393 (1998)
\bibitem{G} L. Ya. Glozman, Surveys in High Energy Physics - in print,
hep-ph/9805345.
\bibitem{RB} D.O. Riska, G.E. Brown, hep-ph/9902319. 
\bibitem{GPP96} L. Ya. Glozman, Z. Papp and W. Plessas, Phys. Lett. {\bf
381}, 311 (1996).
\bibitem{HAR} M. Harvey, Nucl Phys. {\bf A352}, 351 (1981).
\bibitem{BS98} D. Bartz and Fl. Stancu, Phys. Rev. {\bf C59}, 1756 (1998).
\bibitem{NST} V.G. Neudatchin, Yu.F. Smirnov, and R. Tamagaki,
Progr. Theor. Phys. {\bf 58}, 1072 (1977).
\bibitem{OY} M. Oka and K. Yazaki, in Quarks and Nuclei,
International Review of Nuclear Physics, vol. 1, ed. W. Weise
(World Scientific, Singapore, 1984), p. 490
\bibitem{OB} I.T. Obukhovsky and A.M. Kusainov, Yad. Fiz. {\bf 47}, 494
(1988) [Sov. J. Nucl. Phys. {\bf 47}, 313 (1988)].
 \bibitem{GLBK} L.Ya. Glozman, N.A. Burkova and E.I. Kuchina, Z. Phys. {\bf
A332}, 339 (1989).
\bibitem{GLKU} L.Ya. Glozman, V.G. Neudatchin and I.T. Obukhovsky, Phys.
Rev. {\bf C48}, 389 (1993); L Ya. Glozman and E. I. Kuchina,
Phys. Rev. {\bf C49}, 1149 (1994).
\bibitem{SG} K. Shimizu, L.Ya. Glozman, nucl-th/9906008 
\bibitem{ST96} Fl.Stancu {\it Group Theory in Subnuclear Physics},
Clarendon Press, Oxford, 1996, ch. 10
\bibitem{HECHT} K. T. Hecht, Nucl. Phys. {\bf 62}, 1 (1965)
\end{references}
\end{document}